\documentclass[letterpaper,12pt]{extarticle}
\usepackage{preamble}
\usepackage{packages}
\usepackage{breqn}
\usepackage{enumerate}
\usepackage[font=small,labelfont=bf]{caption}
\usepackage{subcaption}
\usepackage[left=30mm,right=30mm,top=30mm,bottom=30mm]{geometry}
\usepackage{pgfplots}
\usetikzlibrary{calc}
\usepackage{extarrows}
\definecolor{refkey}{rgb}{0.9451,0.2706,0.4941}
\definecolor{labelkey}{rgb}{0.9451,0.2706,0.4941}

\definecolor{darkgray}{rgb}{0.33, 0.33, 0.33}
\usepackage{braket}
\usepackage{authblk}
\allowdisplaybreaks

\usepackage{newcommands}

\title{{Stability Analysis of a Non-Unitary CFT}}
\author{Masataka Watanabe\footnote{max.washton@gmail.com}}
\affil{{\small Department of Particle Physics and Astrophysics,\\
Weizmann Institute of Science, Rehovot 7610001, Israel}}
\date{}

\begin{document}
\maketitle

\begin{abstract}
\centering\begin{minipage}{0.86\textwidth}
We study instability of the lowest dimension operator (\it i.e., \rm the imaginary part of its operator dimension) in the rank-$Q$ traceless symmetric representation of the $O(N)$ Wilson-Fisher fixed point in $D=4+\epsilon$.
We find a new semi-classical bounce solution, which gives an imaginary part to the operator dimension of order $O\left({{{\epsilon^{-1/2}}}}\exp\left[-\frac{N+8}{3\epsilon}F(\epsilon Q)\right]\right)$ in the double-scaling limit where $\epsilon Q \leq \frac{N+8}{6\sqrt{3}}$ is fixed.
The form of $F(\epsilon Q)$, normalised as $F(0)=1$, is also computed.
This non-perturbative correction continues to give the leading effect even when $Q$ is finite, indicating the instability of operators for any values of $Q$.
We also observe a phase transition at $\epsilon Q=\frac{N+8}{6\sqrt{3}}$ associated with the condensation of bounces, similar to the Gross-Witten-Wadia transition.






\end{minipage}
\end{abstract}
\newpage
\pagenumbering{arabic}
\pagestyle{plain}
\tableofcontents
\newpage

\section{Introduction}
\label{sec:introduction}




Conformal field theories (CFTs) describe various interesting phenomena in physics even when they lack unitarity.
Even though it is impossible to realise them in a closed system in an equilibrium, they can be realised in experiments as dynamical quantum phase transitions of closed as well as open systems \cite{Ashida:2020dkc}.
One famous example of non-unitary CFTs is the Lee-Yang fixed point in two dimensions \cite{Yang:1952be,Lee:1952ig,Cardy:1985yy}, which is also experimentally realised in real-time dynamics \cite{PhysRevLett.110.050601,PhysRevLett.118.180601}.

Another closely related example of a non-unitary CFT is the $O(N)$ Wilson-Fisher fixed point in $4<D<6$.
This model can either be studied at large-$N$ or by using the $\epsilon$-expansion.
For the latter, one can realise the fixed point as an infrared (IR) fixed point of the Lagrangian with cubic interactions, or as a formal ultraviolet (UV) fixed point of the Lagrangian with a quartic potential.

Non-unitary as it is, the model shows non-unitarity in a subtle way.
Let us imagine computing the operator dimension of $\phi$, the scalar field in the fundamental representation.
In the large-$N$ expansion, the result is real and positive and hence compatible with unitarity to all orders in $1/N$.
Only after we take the instanton corrections into account, does the operator dimension gets an imaginary part of order $O(e^{-N})$, meaning that the theory is non-unitary \cite{Giombi:2019upv}.

The situation is the same in the $\epsilon$-expansion.
In $D=6-\epsilon$, one can show that all the coupling constants at the fixed point are real for large enough $N>N_{\rm crit}$, so that the operator dimension is again real and positive to all orders in the perturbation theory \cite{Fei:2014yja}.
However, due to the cubic coupling, the vacuum at $\phi=0$ can tunnel off to infinity, and hence the operator dimension gets a non-perturbative imaginary part of order $O(e^{-1/\epsilon})$ \cite{Giombi:2019upv}.
In $D=4+\epsilon$, likewise, the negative sign of the quartic coupling at the fixed point is responsible for the non-perturbative imaginary part of the operator dimension \cite{McKane:1984eqmod}.


Even though this was an indirect evidence that the theory is non-unitary, one can see its non-unitary nature more directly by going to the sector of large representations.
As an object of interest, we will take the lowest dimension operator $\mathcal{O}_Q$ in the rank-$Q$ traceless symmetric representation, and we will denote its dimension as $\Delta(Q)$.\footnote{For attempts to generalise this to other representations, see \cite{Hellerman:2017efx,Hellerman:2018sjf,Banerjee:2019jpw,Banerjee:2021bbw}.}
In particular, we are interested in the limit where $Q\to \infty$.

We will now review the generalities regarding such operators in $D$-dimensional $O(N)$ Wilson-Fisher fixed point, putting aside the issue of non-unitarity for the moment.
We denote the scalar fields in the fundamental representation as $\phi^a$, and we will define $\varphi^i\equiv \left(\phi^{a=2i-1}+i\phi^{a=2i}\right)/\sqrt{2}$ for later convenience.
First of all, at large-$Q$, $\mathcal{O}_Q$ are described by an effective field theory (EFT) around a semi-classical saddle-point configuration \cite{Hellerman:2015nra,Alvarez-Gaume:2016vff} (For more references see \cite{Gaume:2020bmp} and references therein.) of the form
\begin{align}
    \begin{cases}
        \varphi_1=\frac{a}{\sqrt{2}}\, e^{i\omega t}\\
        \varphi_i=0 & (i\neq 1)
    \end{cases}.
\end{align}
The EFT, which we will not reproduce here, tells us that the leading order behaviour of $\Delta(Q)$ becomes
\begin{align}
    \Delta(Q)=c_0Q^{D/(D-1)}+O\left(Q^{({D-2})/({D-1})}\right).
\end{align}
Since this is a semi-classical analysis, we could also have guessed this from the dimensional analysis.
Note that $c_0$ is an unknown coefficient appearing in the Wilsonian effective action, and hence we cannot determine it without using the UV data.

Second, we can combine the analysis with the $\epsilon$-expansion or the large-$N$ expansion, where there is a nice double-scaling limit which fixes either $\epsilon Q$ or $Q/N$ \cite{Alvarez-Gaume:2019biu,Badel:2019oxl,Watanabe:2019pdh,Grassi:2019txd}.
This computation can be done by noticing that $a$ and $\omega$ above can be computed by using the equation of motion (EOM) and the charge fixing constraint, and that this is a controlled approximation at weak coupling.

If we take the $D=4-\epsilon$ case as an example, the energy of this configuration is of the form $\frac{1}{\epsilon}F_0(\epsilon Q)$, which, on the unit sphere, is nothing but $\Delta(Q)$ at leading order in $\epsilon$ but fixed $\epsilon Q$.
By doing the loop computations, we can see that the result is of the form
\begin{align}
    \Delta(Q)=\frac{1}{\epsilon}F_0(\epsilon Q)+F_1(\epsilon Q)+\epsilon {F_2(\epsilon Q)}+\cdots
\end{align}
and hence we find a double-scaling limit which fixes $\epsilon Q$.
Furthermore, we find that
\begin{align}
    F_0(\epsilon Q)=
    \begin{cases}
        \epsilon Q+O(\epsilon Q^2) & \text{when $\epsilon Q\ll 1$}\\
        b_1(\epsilon Q)^{D/(D-1)}+O\left((\epsilon Q)^{(D-2)/(D-1)}\right) & \text{when $\epsilon Q\gg 1$}
    \end{cases}
    \label{eq:preresult}
\end{align}
as expected from the weak-coupling intuition and from the EFT intuition at $\epsilon Q\ll 1$ and at $\epsilon Q \gg 1$, respectively.
Here, $b_1$ does not depend on $Q$ nor $\epsilon$, and is just a real number which can be computed in principle.
The same story goes for the large-$N$ expansion, where the double-scaling parameter is $Q/N$ \cite{Alvarez-Gaume:2019biu}.

Let us now go back to the original model we are interested in, the $O(N)$ Wilson-Fisher fixed point in $4<D<6$.
The model also has a solvable limit around $D=4,\, 6$ or at large-$N$, and hence the above double-scaling limit is still applicable.
Let us for the moment concentrate on the $D=4+\epsilon$ case.
There is a big difference which distinguishes the current case from the unitary case -- 
when $\epsilon Q$ is bigger than $\lambda_0\equiv \frac{N+8}{6\sqrt{3}}$, the real saddle-point for $\phi^a$ ceases to exist, as discussed in \cite{Giombi:2020enj,Antipin:2021jiw}.\footnote{In fact, \cite{Giombi:2020enj} saw this at large $N$ and \cite{Antipin:2021jiw} did the same in $D=6-\epsilon$. However, the current discussion is a trivial generalisation of those cases.}
The energy of this complex configuration therefore becomes complex as well.\footnote{If we study the same model in $D=4-\epsilon$, the point $\epsilon Q=-\lambda_0$ renders the convergence radius of the small $\epsilon Q$ expansion to a finite value, $\abs{\epsilon Q}=\lambda_0$. The phenomenon that the perturbative expansion breaks down at some finite value of $\epsilon Q$ and is taken over by a different semi-classical description is also observed in the multiparticle cross section of scalar theories \cite{Libanov:1994ug,Libanov:1995gh,Son:1995wz,Bezrukov:1995qh}.}
This can also be seen immediately by plugging in $-\epsilon$ into $\epsilon$ in \eqref{eq:preresult}.
All in all, it is now easy to see that our model is non-unitary, since the operator dimension of $\mathcal{O}_Q$ has a huge imaginary part at large-$Q$.

Having said that, what we are interested in in this paper is the regime where $\epsilon Q\leq\lambda_0$.
In this regime, the saddle-point configuration is real, and the operator dimension which follows from it is also real to all orders in $\epsilon Q$ and in $\epsilon$.
Furthermore, the result correctly reproduces (up to the order where computations on both sides are done anyway, but presumably to all orders) the Feynman diagram computation if we plug in some fixed value of $Q$.
On the other hand, we know that such a computation, even if it were done to all orders, will not reproduce the non-perturbative imaginary part of the operator dimension observed at finite $Q$, by definition.
Nevertheless, it still should be visible in the double-scaling limit as this non-perturbative correction gives the leading part in the imaginary part of the operator dimension.

In this paper, by conducting a more careful saddle-point analysis of the $O(N)$ Wilson-Fisher fixed point in $D=4+\epsilon$, we find this non-perturbative correction in the double-scaling limit.
We see that this is due to a time-dependent semi-classical saddle in the Euclidean path-integral, called the bounce solution.
By computing the action of this bounce and conducting a one-loop computation around it, the leading contribution to the imaginary part of $\Delta(Q)$ is computed as
\begin{align}
    \im \Delta(Q)=\pm \sqrt{\frac{2(N+8)}{\pi\epsilon}}f_b(\epsilon Q)\exp\left(-\frac{N+8}{3\epsilon}F(\epsilon Q)\right)
\end{align}
where $F(\epsilon Q)$ is some function which satisfies $F(0)=1$ and $F(\lambda_0)=0$, while $f(\epsilon Q)$ is also some function which satisfies $f(0)=1$.
These functions are computed in the main text.

We find two interesting phenomena regarding this.
First, the exponent $-\frac{N+8}{3\epsilon}F(\epsilon Q)$ is expected to persist even at finite $Q$, since further corrections in the double-scaling limit will not interfere with this number even at finite $Q$.
Therefore, we expect that the leading imaginary part of the operator dimension at fixed $Q$ is $O\left(\exp\left(-\frac{N+8}{3\epsilon}\right)\right)$.
This correctly reproduces the non-perturbative imaginary part of $\Delta(Q)$ computed in a conventional way in \cite{McKane:1984eqmod}.

Second, $F(\epsilon Q)$ is a decreasing function and hits zero at $\epsilon Q=\lambda_0$.
As we increase $\epsilon Q$, the imaginary parts of $\Delta(Q)$ gets bigger and bigger, and eventually at $\epsilon Q=\lambda_0$, it becomes $O(1)$.
Therefore the dilute-gas approximation that we implicitly used is not to be trusted anymore.
This is the bounce version of the so-called instanton condensation, observed, for example, in the Gross-Witten-Wadia model \cite{Wadia:2012fr,Wadia:1980cp,Gross:1980he,Gross:1994mr,Gross:1994ub,Witten:1991we,Marino:2008ya}.\footnote{A similar phenomenon is also discussed in the context of the Mathieu equation \cite{Basar:2015xna,Dunne:2016qix}.}
In fact, as in the Gross-Witten-Wadia transition, this condensation reflects the fact that the function $F_0(\epsilon Q)$ shows a phase transition at $\epsilon Q=\lambda_0$.
This is in contrast to the $D=4-\epsilon$ case, where the function $F_0(\epsilon Q)$ just shows a crossover at around $\epsilon Q=\lambda_0$ and correspondingly there are no non-perturbative saddles which show the condensation.

The rest of the paper is organised as follows.
In Section \ref{sec:onwf}, we review the $O(N)$ Wilson-Fisher fixed point in $4<D<6$ and its double-scaling limit which probes the lowest operator dimension in rank-$Q$ traceless symmetric representations.
In Section \ref{sec:operatordimension}, we compute the non-perturbative imaginary part of the operator dimension which comes from the bounce when $0<\epsilon Q<\lambda_0$.
In Section \ref{sec:discussion}, we look more closely at two interesting points -- at $\epsilon Q=0$ we reproduce the size of the non-perturbative imaginary part of the operator dimension at finite $Q$, and at $\epsilon Q=\lambda_0$ we find the condensation of bounces, indicating the breakdown of the dilute-gas approximation. 
Finally, we summarise our results and give possible future directions in Section \ref{sec:conclusion}.


\section{$O(N)$ Wilson-Fisher fixed point in $4<D<6$}
\label{sec:onwf}

\subsection{The model}

The model we are going to study is the $O(N)$ Wilson-Fisher theory in $4<D<6$.
Unlike its counterpart in $2<D<4$, this model is peculiar in that it can never truly be unitary.
This model is usually studied at large-$N$ or in the $\epsilon$-expansion, and in this paper we mostly consider the latter.
For more through discussion of the model and its beta functions, the readers are referred to \cite{Fei:2014yja,Fei:2014xta,Giombi:2019upv}.

\subsubsection{$4+\epsilon$ dimensions}

In $D=4+\epsilon$, the theory can be realised as a formal UV fixed point of the Lagrangian with the quartic potential in $D=4+\epsilon$,
\begin{align}
    L=\frac{1}{2}(\partial \phi^a)^2+\frac{m^2}{2}(\phi^a \phi^a)+\frac{g}{4}(\phi^a \phi^a)^2
\end{align}
Here, $\phi^a$ is in the fundamental representation of $O(N)$.
Additionally, $m$ is the conformal coupling, and is $m=(D-2)/2$ on $S^{D-1}\times \mathbb{R}$.

The beta function and the coupling constant at the fixed point of the Lagrangian are readily available in the literature \cite{Giombi:2020enj},
\begin{align}
    \beta(g)=\epsilon g+\frac{N+8}{8\pi^2}g^2 +O(g^3), \quad g_{*}=-\frac{8\pi^2\epsilon}{N+8}+O(\epsilon^2)
    \label{eq:fixed-point-coupling}
\end{align}
at leading order in $\epsilon$, for any $N$.

The fact that $g_{*}$ is negative indicates that the theory is unstable as discussed in \cite{Giombi:2020enj}.
Therefore, non-perturbative corrections from instantons gives the coupling constant, as well as to the operator dimensions, imaginary parts of order $O(e^{-1/\epsilon})$.
However, the perturbative part of the coupling constant stays real for all values of $N\geq 2$.

\subsubsection{$6-\epsilon$ dimensions}

In $D=6-\epsilon$, the theory is realised as the IR fixed point of the Lagrangian with the cubic potential,
\begin{align}
    L=\frac{1}{2}(\partial \phi^a)^2+\frac{m^2}{2}(\phi^a \phi^a)+\frac{1}{2}(\partial \sigma)^2+\frac{m^2}{2}\sigma^2+\frac{p}{2}\sigma\phi^a \phi^a+\frac{q}{6}\sigma^3.
\end{align}
Here, $\phi^a$ is in the fundamental representation of $O(N)$, while $\sigma$ is a singlet under it.

The beta functions are known to be \cite{Giombi:2020enj}
\begin{align}
    \begin{split}
        \beta(p)&=-\frac{\epsilon}{2}p+\frac{(N-8)p^3-12p^2 q + p q^2}{12(4\pi)^3}\\
        \beta(q)&=-\frac{\epsilon}{2}q+ \frac{-4Np^3 +Np^2 q -3q^3}{4(4\pi)^3}
    \end{split}.
\end{align}
The coupling constants at the fixed point are real as long as $N>N_{\rm crit}=1038.3\dots$,
where in the $1/N$-expansion,
\begin{align}
    \begin{split}
        p_* &= \sqrt{\frac{6\epsilon(4\pi)^3}{N}}\left( 1 + \frac{22}{N}+\frac{726}{N^2}-\frac{326180}{N^3}  +\cdots  \right)+O\left(\epsilon^{3/2} \right)  \\
        q_* &= 6 \sqrt{\frac{6\epsilon(4\pi)^3}{N}}\left(1 + \frac{162}{N}+\frac{68766}{N^2}+\frac{41224420}{N^3}\cdots\right)+ O\left(\epsilon^{3/2} \right)
    \end{split}.
\end{align}
Since the Lagrangian has the cubic potential, the perturbatively real coupling constants at the fixed point get non-perturbatively corrected and obtain imaginary parts even when $N>N_{\rm crit}$.
On the other hand, when $N<N_{\rm crit}$, the perturbative part of the coupling constant is already imaginary.

\subsection{Partition function in the sector of fixed charge}

We are interested in the lowest operator dimension of the traceless symmetric representation of the $O(N)$ model of degree $Q$.
Since we are considering complex CFTs, by lowest we mean that it has the smallest real parts.
One easy way to compute this is to use the state-operator correspondence and map the problem to the ground state energy of this particular sector.

For simplicity, we take $N$ to be an even number hereafter.
We then package $N$ number of real fields $\phi^a$ into $N/2$ complex fields, $\varphi^i\equiv \left(\phi^{a=2i-1}+i\phi^{a=2i}\right)/\sqrt{2}$.
We denote the charge rotating the field $\varphi^i$ to be $\hat{Q}_i$, which is defined as
\begin{align}
    \hat{Q}_i\equiv i\int_{S^{D-1}} d\vec{x}\left[\partial_0\varphi_i^*\varphi_i-\partial_0\varphi_i\varphi_i^*\right].
\end{align}
We also denote the charge densities by $\rho_i$.
Looking for the lowest operator dimension in the traceless symmetric representation of degree $Q$ is equivalent to looking for the lowest operator dimension in the sector where $\sum_{i=1}^{N/2} Q_i=Q$.
This is because the states in asymmetric representations have more energy than the symmetric ones in our case \cite{Alvarez-Gaume:2016vff,Hellerman:2017efx,Hellerman:2018sjf}.

To compute the ground state energy in such a sector, we use the (projected) canonical partition function \cite{Alvarez-Gaume:2019biu},
\begin{align}
    \begin{split}
        Z(\beta,Q)\equiv \Tr\left[e^{-\beta H}
        \delta(\hat{Q}-Q)\right]=\int_{-\pi}^{\pi}
        \left\{
        \frac{d\theta}{2\pi}e^{i\theta Q}
        \right\}
        \Tr\left[e^{-\beta H-i
        \theta \hat{Q}}\right].
    \end{split}
    \label{eq:projectedpf}
\end{align}
Incidentally, this is a special case of the character decomposition, which should be used for general representations other than the symmetric one.\footnote{However, the technicality could be quite cumbersome. We thank Domenico Orlando for discussions.}
By taking $\beta\to\infty$, we extract the information about the ground state energy, which is nothing but the lowest operator dimension in the rank-$Q$ traceless symmetric representation
\begin{align}
    \Delta(Q)=-\lim_{\beta\to\infty} \frac{1}{\beta}\log Z(\beta,Q).
\end{align}

Meanwhile, the integrand in \eqref{eq:projectedpf} is nothing but the grand canonical partition function with imaginary chemical potential.
This can be represented as a Euclidean path integral
\begin{align}
    \Tr\left[e^{-\beta H-i\sum_i \theta_i \hat{Q}_i}\right]
    =\int_{\text{b.c.}}\mathcal{D}\varphi\, e^{-S[\varphi]},
    \label{eq:twistedboundarypathintegral}
\end{align}
where we impose the twisted boundary condition such that
\begin{align}
    \varphi_i(\beta,\vec{x})=e^{i\theta}\varphi_i(0,\vec{x}).
\end{align}
Additionally, the integral over $\theta$ is typically dominated by a saddle-point at large $\beta$.
This reduces the computation of \eqref{eq:projectedpf} to the ``Legendre transformation'' of the the grand canonical partition function with imaginary chemical potential.

One caveat, which is why we put ``Legendre transformation'' into the quotation mark, is that the grand canonical partition function is not really a convex function for $4<D<6$ where the theory is not unitary \cite{Moser:2021bes,Orlando:2021usz}.\footnote{Moreover, for free theories or for theories with moduli, the chemical potential gets fixed to the free theory value and ``Legendre transformation'' becomes misnomer as well \cite{Hellerman:2017veg,Hellerman:2017sur,Hellerman:2018xpi,Hellerman:2020sqj,Sharon:2020mjs}.}
Indeed, this is related to the fact that the value of the chemical potential, $\omega\equiv i\theta/\beta$, at the saddle-point develops an imaginary part, as we explain later.
When such is the case, a saddle-point in terms of $\theta$ is always paired with its complex conjugate, and hence there will be two lowest dimension operator (in the sense that the real part is the lowest) in the representation in question.

\subsection{Saddle-point equation and the double-scaling limit}

\subsubsection{Equation of motion and charge fixing constraint}

Let us now evaluate \eqref{eq:projectedpf} using the saddle-point approximation at large $Q$ and small $\epsilon$.
The saddle-point equation for \eqref{eq:twistedboundarypathintegral} is nothing but the equation of motion, which is
\begin{align}
    \partial^\mu\partial_\mu \varphi_i=m^2\varphi_i+2g\left(\sum_{i=1}^{N/2}\abs{\varphi_j}^2\right)\varphi_i.
    \label{eq:EOM}
\end{align}
Furthermore, we impose the charge fixing constraint,
\begin{align}
    Q= i\int_{S^{D-1}} d\vec{x}\left[\sum_{i=1}^{N/2}\left(\partial_0\varphi_i^*\varphi_i-\partial_0\varphi_i\varphi_i^*\right)\right].
\end{align}
This comes from the saddle-point equation of the $\theta$-integral in \eqref{eq:projectedpf}, where the saddle-point approximation is exact in the strict $\beta\to\infty$ limit we are interested in.


We now look for the lowest energy solution to this equation of motion (EOM) subject to the charge fixing constraint.
Such a solution is homogeneous in space and helical in time.
As discussed in \cite{Gaume:2016vff,Antipin:2020abu,Gaume:2020bmp}, by using the symmetry of the problem we can set the configuration to be
\begin{align}
    \begin{cases}
        \varphi_1=\frac{a}{\sqrt{2}}\, e^{i\omega t}\\
        \varphi_i=0 & (i\neq 1)
    \end{cases}.
\end{align}
The EOM and the charge fixing constraint therefore reduce to
\begin{align}
    a^2=-\frac{\omega^2-m^2}{\abs{g}},\quad
    Q=\omega a^2\alpha(D),
    \label{eq:eom+cfc}
\end{align}
where $\alpha(D)$ is the area of the unit $(D-1)$-dimensional sphere,
\begin{align}
    \alpha(D)=\frac{2\pi^{D/2}}{\Gamma(D/2)}.
\end{align}

\subsubsection{The double-scaling limit and the saddle-point}
\label{sec:staticeom}

The coupled equations in \eqref{eq:eom+cfc} reduces to a cubic equation for $\omega$,
\begin{align}
    \omega^3-m^2\omega=-\frac{\abs{g}Q}{\alpha(D)}\equiv -\gamma.
    \label{eq:cubic}
\end{align}
It is now apparent that there is a double-scaling limit which fixes $gQ$ since the saddle-point only depends on the combination $gQ$.

Let us first find all the solutions for \eqref{eq:cubic}, and then think about their physical meaning later on.
Since this is a cubic equation, there are three solutions $\omega_{\pm,n}$ all of which can be expressed analytically,
\begin{align}
    \omega_0\equiv \omega_+&=\frac{2^{1/3} \left(-9 \gamma +\sqrt{81 \gamma ^2-12 m^6}\right)^{2/3}+2\cdot 3^{1/3} m^2}{6^{2/3} \left({-9 \gamma +\sqrt{81 \gamma ^2-12 m^6}}\right)^{1/3}}\\
    &=m\left(1-\frac{\gamma}{2m^3}-\frac{3\gamma^2}{8m^6}+O(\gamma^3)\right)\\
    \omega_-&=-\frac{e^{i\pi/3}2^{1/3} \left(-9 \gamma +\sqrt{81 \gamma ^2-12 m^6}\right)^{2/3}+e^{-i\pi/3}2\cdot 3^{1/3} m^2}{6^{2/3} \left({-9 \gamma +\sqrt{81 \gamma ^2-12 m^6}}\right)^{1/3}}\\
    \omega_n&=-\frac{e^{i2\pi/3}2^{1/3} \left(-9 \gamma +\sqrt{81 \gamma ^2-12 m^6}\right)^{2/3}-e^{i\pi/3}2\cdot 3^{1/3} m^2}{6^{2/3} \left({-9 \gamma +\sqrt{81 \gamma ^2-12 m^6}}\right)^{1/3}}
\end{align}
When $\gamma>0$, $\omega_n$ is always real and negative.
For the other two solutions, the situation is more complex.
We have that when $0<\gamma<\gamma_0$, we have $0<\omega_-<\omega_+$, while when $\gamma>\gamma_0$, they are complex and not real anymore, with one being the complex conjugate of the other, where
\begin{align}
    \gamma_0\equiv \frac{2m^3}{3\sqrt{3}}.
\end{align}
When $\gamma=\gamma_0$, $\omega_{\pm}$ coincides with each other.
This situation is parallel to what is discussed in \cite{Giombi:2020enj,Antipin:2021jiw}.

We now need to understand which of these three saddle-points are relevant for computing the lowest operator dimension.
We first rule out $\omega_n$ for being the local maximum of the action for any values of $\gamma$.
Out of the remaining two, $\omega_+$ is the physically relevant saddle-point -- when $\gamma<\gamma_0$, this gives the local minimum of the action while $\omega_-$ gives the local maximum.
Furthermore, the configuration with $\omega_+$ has the lowest energy than that with $\omega_-$ when $\gamma<\gamma_0$, which is another reason why we favour $\omega_+$ over $\omega_-$.
Meanwhile, when $\gamma>\gamma_0$, $\omega_{\pm}$ is complex conjugate to each other, so both of them are physically relevant.
Since the resulting operator dimensions are also complex conjugate to each other, without loss of generality, we will only care about the saddle-point at $\omega=\omega_+$ hereafter.
We will define $\omega_0\equiv \omega_+$, and also denote the resulting value for $a$ as $a_0$ as well.

\subsubsection{The lowest operator dimension}
\label{sec:phasetransition}

The energy of the saddle-point configuration gives us the lowest operator dimension in the rank-$Q$ traceless symmetric representation.
In our double-scaling limit, this is schematically given by
\begin{align}
    \Delta(Q)=\frac{1}{\epsilon}F_0(\epsilon Q)+F_1(\epsilon Q)+\epsilon {F_2(\epsilon Q)}+\cdots+\text{(non-perturbative)}
\end{align}
We denote the perturbative part of this expression as $\Delta_{\rm pert}(Q)$.

\begin{figure}[tbp]
    \begin{subfigure}[htbp]{0.5\columnwidth}
        \begin{center}
            \begin{overpic}[ width=0.97\columnwidth]{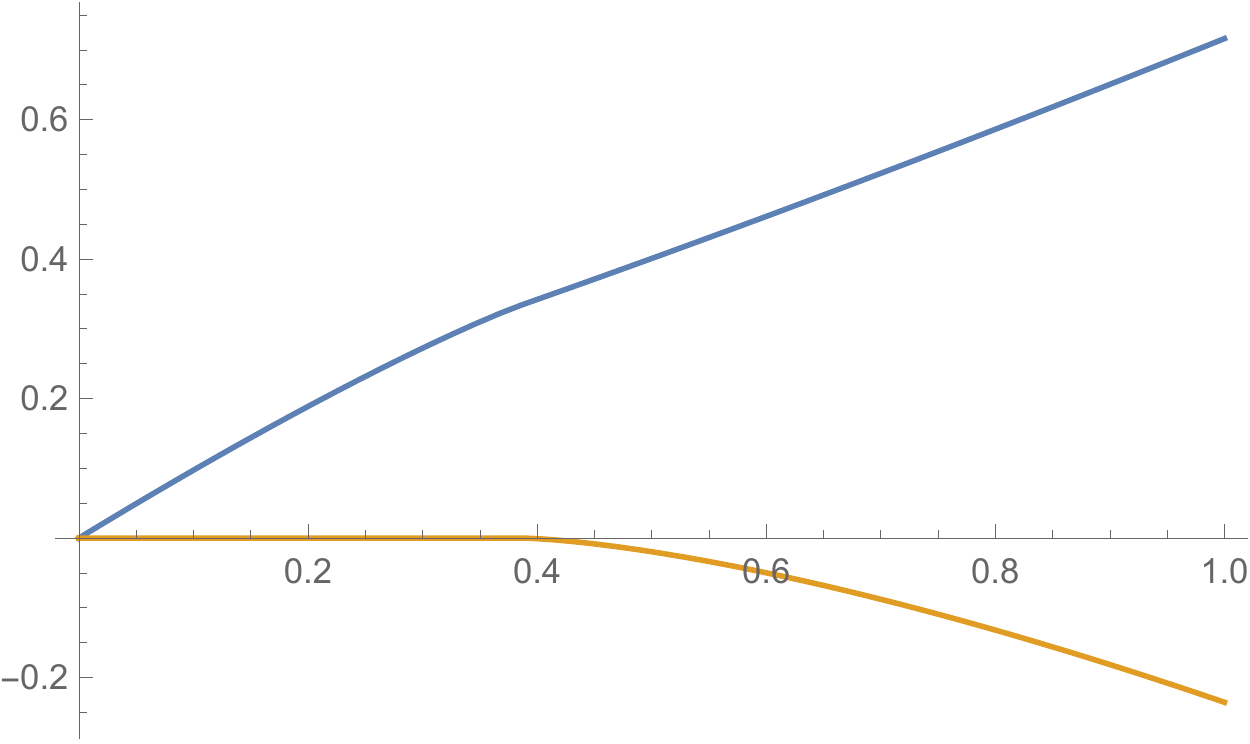}
            \put(41,18){\footnotesize $\gamma_0$}
            \put(3,62){\footnotesize $\tilde{F}_0(\gamma)$}
            \end{overpic}
        \end{center}
    \end{subfigure}
    \hfill
    \begin{subfigure}[htbp]{0.5\columnwidth}
        \begin{center}
            \begin{overpic}[width=0.97\columnwidth]{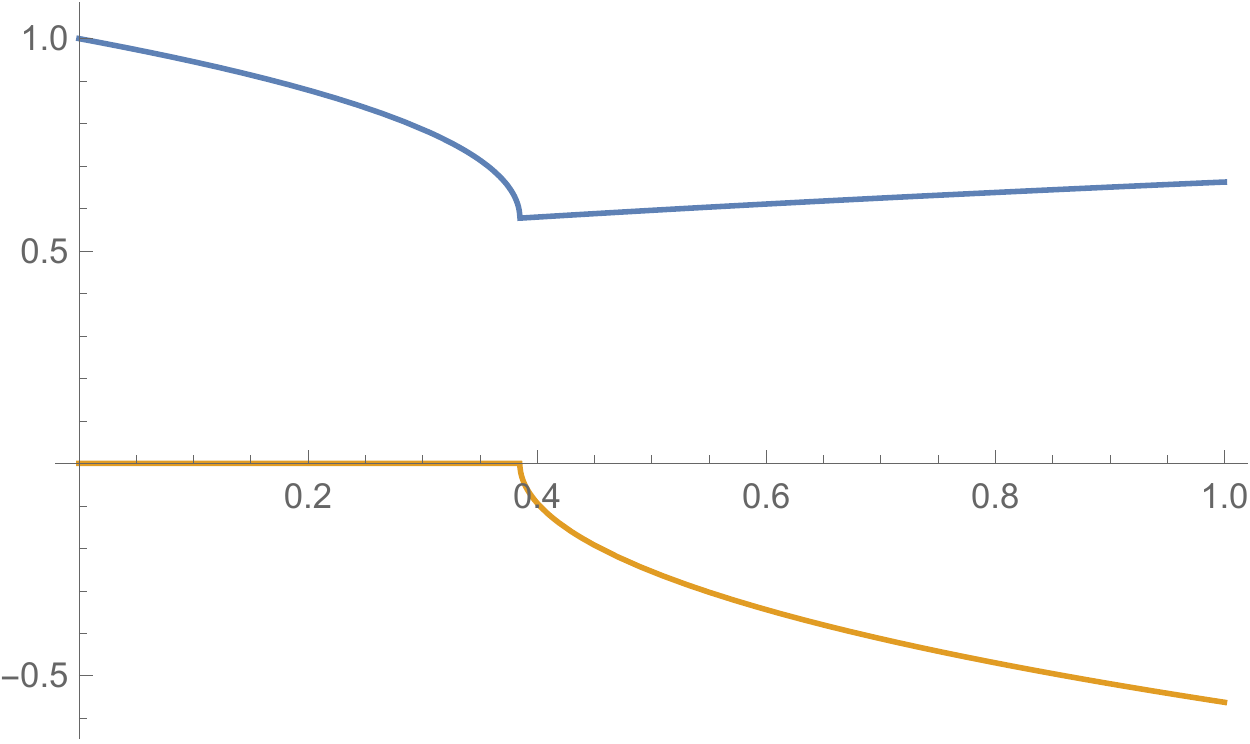}
            \put(39,24){\footnotesize $\gamma_0$}
            \put(3,62){\footnotesize $\tilde{F}_0^{\prime}(\gamma)$}
            \end{overpic}
            \end{center}
    \end{subfigure}
    \caption{The shape of the function $\tilde{F}_0(\gamma)$ and its first derivative, $\tilde{F}_0^{\prime}(\gamma)$. The blue line indicates the real part, while the orange indicates the imaginary part. {\bf Left}: The function $\tilde{F}_0(\gamma)$, which is a smooth function in terms of $\gamma$. 
    For simplicity, we have put $m=1$ to the expression.
    {\bf Right}: The first derivative of $\tilde{F}_0(\gamma)$. We see that it develops a kink at $\gamma=\gamma_0$, indicating a second-order phase transition. We also see that the real part of $\tilde{F}_0(\gamma)$ is a concave function for $0<\gamma<\gamma_0$, while convex for $\gamma>\gamma_0$.}
    \label{fig:f0}
\end{figure}

For later reference, let us compute $\frac{1}{\epsilon}F_0(\epsilon Q)$.
We have
\begin{align}
    \frac{1}{\epsilon}F_0(\epsilon Q)=\frac{3^{1/3} \alpha(D)}{2^{10/7}\abs{g}}\tilde{F}_0(\epsilon Q)
    \label{eq:leading}
\end{align}
where
\begin{align}
    \tilde{F}_0(\epsilon Q)\equiv\frac{\gamma\left(A(\gamma)^{4/3}+2^{2/3}\cdot3^{4/3}m^2A(\gamma)^{2/3}+3^{2/3}\cdot2^{4/3}m^4\right)}{A(\gamma)+2^{2/3}\cdot3^{1/3}A(\gamma)^{1/3}}
\end{align}
and 
\begin{align}
    A(\gamma)\equiv -9\gamma+\sqrt{81\gamma^2-12m^6}.
\end{align}
We depicted the form of $\tilde{F}_0(\gamma)$ in Fig. \ref{fig:f0}.
Importantly, this function shows a second-order phase transition at $\gamma=\gamma_0$.
We also expect that this continues to be the case for $\Delta_{\rm pert}(Q)$ as well.
This is to be contrasted with the $D=4-\epsilon$ case discussed in \cite{Badel:2019oxl,Watanabe:2019pdh}, where $\Delta_{\rm pert}(\epsilon Q)$ can be differentiated as many times as one wants for any $\epsilon Q$. 
This phase transition will be discussed later on in Sec. \ref{sec:instantoninduced} as well as its relation to the condensation of bounces.

Since what we are mostly interested in is the regime where $0<\gamma<\gamma_0$, let us also compute $\Delta(Q)$ when $\epsilon Q\ll 1$.
This is already computed in the literature \cite{Badel:2019oxl,Watanabe:2019pdh} -- up to one-loop, $\Delta(Q)$ can be expanded at small $gQ$ as
\begin{align}
    \begin{split}
        \Delta_{\rm pert}(Q)=Q\biggl[m&+\frac{g}{8\pi^2}(Q-1)\\
        &\quad+\frac{g^2}{(8\pi^2)^2}\left(2Q^2+\frac{(N-22)(N+6)Q}{N+8}+O(Q^0)\right)+O(\epsilon^3)\biggr].
    \end{split}
\end{align}
Plugging in \eqref{eq:fixed-point-coupling}, we have
\begin{align}
	\begin{split}
		\Delta_{\rm pert}(Q)=Q\biggl[&\left(1+\frac{\epsilon}{2}\right)-\frac{\epsilon (Q-1)}{N+8}\\
		&\quad+\frac{\epsilon^2}{(N+8)^2}\left(2Q^2+\frac{(N-22)(N+6)Q}{N+8}+O(Q^0)\right)+O(\epsilon^3)\biggr].
	\end{split}
\end{align}
As we have already explained, this expression is real to all orders in $\epsilon$ and in $\epsilon Q$.






\section{Non-perturbative instability from the bounce}
\label{sec:operatordimension}

\subsection{Bounce solution in the Euclidean path integral}

\subsubsection{Equation of motion in the Euclidean signature}

Even though $\Delta_{\rm pert}(Q)$ gives us the perturbative and dominant contribution to the operator dimension, there is another solution that contributes to the Euclidean path integral for the projected partition function.
Such a solution must be a time-dependent solution in the Euclidean signature, starting at $\varphi(0,\vec{x})=(a_0,0,\dots, 0)$ and ending at $\varphi(\beta,\vec{x})=(a_0e^{i\omega_E\beta},0,\dots, 0)$ (meaning, sharing the same boundary condition with the lowest helical solution).
Note that $\omega_E$ here is the chemical potential in the Euclidean signature, which is $\omega_E=-i\omega$ while $t_E=it$ is the Euclidean time.

We now look for such a solution.
To avoid various factors of $i$, we first study the EOM for the Lorentzian signature and we will move to the Euclidean signature at the very end.
First of all, because of the charge conservation, the configuration of interest is of the form $\varphi(t,\vec{x})=(\varphi_1(t,\vec{x}),0,\dots, 0)$.
We now separate $\varphi_1$ into radial and angular variables, 
\begin{align}
    \varphi_1=\frac{a}{\sqrt{2}}\, e^{i\chi}.
\end{align}
We also set an ansatz that the solution of interest is homogeneous in space.
After all these, we can rewrite and simplify the EOM \eqref{eq:EOM} as
\begin{align}
    \ddot{a}=a\dot\chi^2-m^2 a-ga^3, \quad a\ddot{\chi}+2\dot{a}\dot{\chi}=0,
\end{align}
where we denoted the time derivative in terms of a dot.

First of all, the second equation is nothing but the charge conservation.
One can see this by integrating it once, obtaining $a^2\dot{\chi}={\rm (const.)}$, where this constant is set by the boundary condition that the charge is $Q$, \it i.e., \rm
\begin{align}
    a^2\dot{\chi}=\frac{Q}{\alpha(D)}.
\end{align}
Note that this is consistent with \eqref{eq:eom+cfc}.

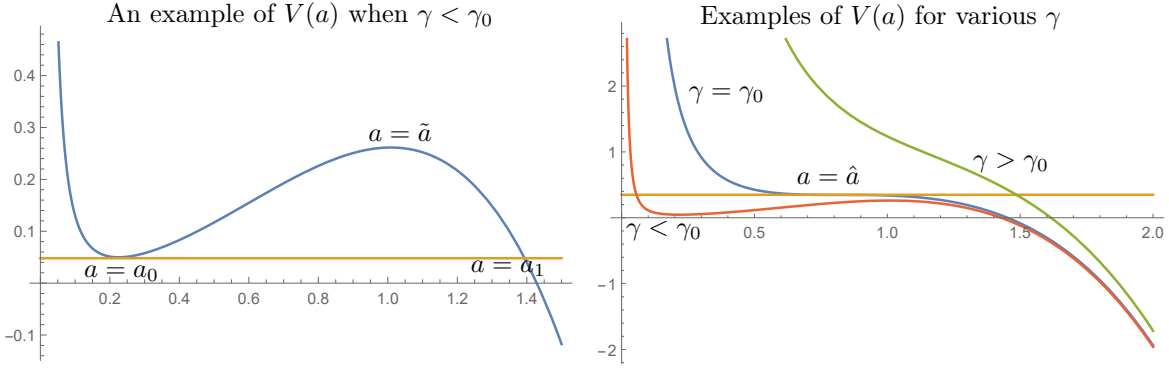
\begin{figure}[tbp]
    \begin{subfigure}[htbp]{0.5\columnwidth}
        \begin{center}
            \begin{overpic}[ width=0.97\columnwidth]{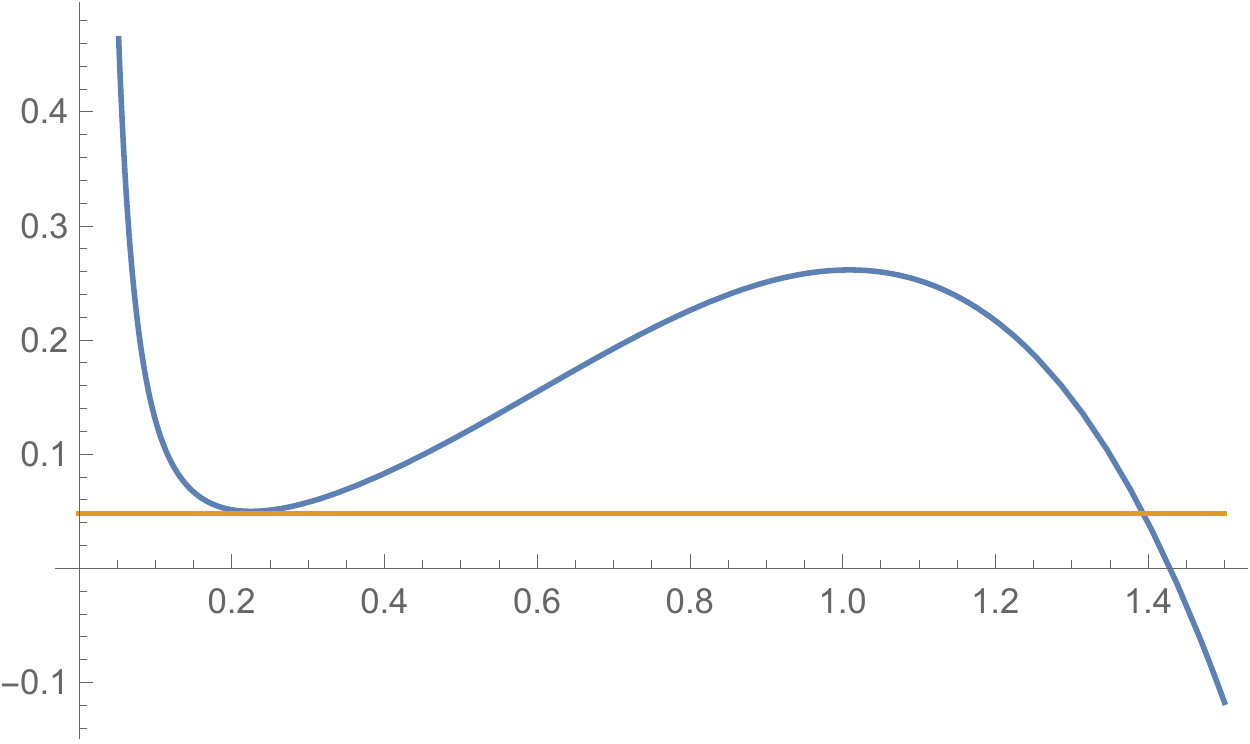}
            \put(14,15){\footnotesize $a=a_0$}
            \put(82,16){\footnotesize $a=a_1$}
            \put(64,39){\footnotesize $a=\tilde{a}$}
            \put(18,60){\footnotesize An example of $V(a)$ when $\gamma<\gamma_0$}
            \end{overpic}
        \end{center}
    \end{subfigure}
    \hfill
    \begin{subfigure}[htbp]{0.5\columnwidth}
        \begin{center}
            \begin{overpic}[width=0.97\columnwidth]{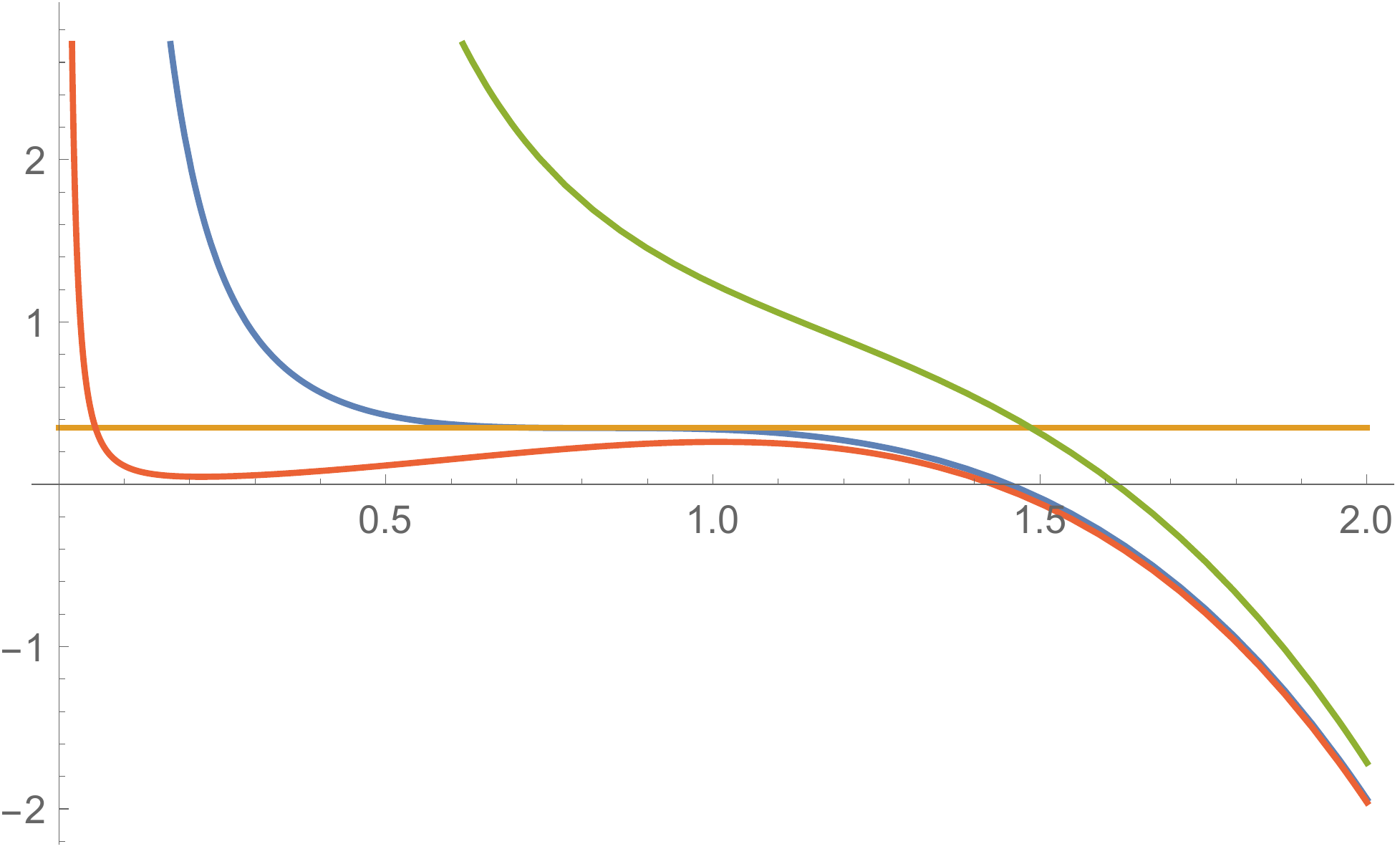}
            \put(35,32){\footnotesize $a=\hat{a}$}
            \put(16,47){\footnotesize $\gamma=\gamma_0$}
            \put(66,35){\footnotesize $\gamma>\gamma_0$}
            \put(5,23){\footnotesize $\gamma<\gamma_0$}
            \put(18,60){\footnotesize Examples of $V(a)$ for various $\gamma$}
            \end{overpic}
            \end{center}
    \end{subfigure}
    \caption{Examples of the potential $V(a)$. The axes are rescaled so that it only depends on $\gamma\equiv \abs{g}\rho$, \it i.e., \rm the horizontal axis is $a\sqrt{\abs{g}}$, while the vertical axis is $\abs{g}V(a)$.
    {\bf Left}: An example of $V(a)$ when $\gamma <\gamma_0\equiv \frac{2m^3}{3\sqrt{3}}$ (blue). The concrete values are taken to be $\gamma=0.05$ and $m=1.01$. $a_{0,1}$ are defined below \eqref{eq:energyconservation}, and the value of $V(a_0)$ is indicated by an orange line. $\tilde{a}$ is irrelevant to the discussion but it is defined as the local maxima of the potential, which is the value of $a$ corresponding to $\omega=\omega_-$ in Sec. \ref{sec:staticeom}.
    {\bf Right}: Example of $V(a)$ from $\gamma<\gamma_0$ (red), passing $\gamma=\gamma_0$ (blue) to $\gamma>\gamma_0$ (green). The concrete values are taken to be $\gamma=\gamma_0-0.35$, $\gamma=\gamma_0$, and $\gamma=\gamma_0+1$, with $m=1.01$.
    We can see that, as we increase $\gamma$, the potential gets shallower and eventually loses its stationary point on the real axis when $\gamma>\gamma_0$. The marginal case where $\gamma=\gamma_0$ has a stationary inflection point, whose value is denoted by $\hat{a}$, while the value of $V(\hat{a})$ is indicated by an orange line.}
    \label{fig:potential}
\end{figure}

Now plugging this back into the first equation, we obtain the differential equation for $a$ alone,
\begin{align}
    \ddot{a}=\frac{\rho^2}{a^3}-m^2 a-ga^3\equiv -V^{\prime}(a)
\end{align}
in the Lorentzian signature, where $\rho\equiv Q/\alpha(D)$.
Integrating over $a$ once, we get the form of the potential
\begin{align}
    V(a)=\frac{\rho^2}{2}\frac{1}{a^2}+\frac{m^2}{2}a^2+\frac{g}{4}a^4.
    \label{eq:potentialstable}
\end{align}
We depict the form of the potential in Fig. \ref{fig:potential}.
Noting that we are interested in the region $0<\gamma<\gamma_0$, there is one local minima and one local maxima of this potential, corresponding to two time-independent solutions (with $\omega>0$) discussed in the previous subsection.
In fact, one can check that the local minima of $V(a)$ is achieved when $a=a_0$.

To get the EOM in the Euclidean signature, one can simply invert the sign of the potential,
\begin{align}
    \ddot{a}= V^{\prime}(a).
\end{align}
Equivalently, by integrating both sides once, we get the conservation equation for energy, which is
\begin{align}
    \frac{\dot{a}^2}{2}-V(a)=-V(a_0)
    \label{eq:energyconservation}
\end{align}
It is now apparent that there is a solution that starts from $a_0$ at $t_E=0$ and rolls down $-V(a)$, hitting $a_1$ and eventually going back to $a_0$ at $t_E=\beta\to\infty$, where $a_1$ was defined by $V(a_1)=V(a_0)$.
This is the bounce solution which contributes to the Euclidean path integral of the projected partition function, at $\beta\to\infty$.
We denote such a solution as $a_b(t_E)$.
Even though the exact form will not be relevant later, we comment that there is an analytic expression for $a_b(t_E)$, which is a complicated function including elliptic functions.

\subsubsection{Action of the bounce}
\label{sec:actionofbounce}

Let us first rewrite the Euclidean action using the new variables $a$.
As we only need to know the classical value of the bounce action for the moment, we will truncate all the fields other than the $s$-wave of $a$.
The truncated Euclidean Lagrangian for $a$ simply becomes
\begin{align}
    L_E[a]=\frac{1}{2}\left(\partial a\right)^2+V(a).
\end{align}
Although $a_b(t_E)$ is defined in $0\leq t_E\leq\beta$, for simplicity we shift its range to $-\beta/2\leq t_E\leq\beta/2$ without changing the physics. 
Moreover, as we take $\beta\to\infty$, we think of $a_b(t_E)$ to be defined on the whole real line of $t_E$.
This means that the bounce solution has the property 
\begin{align}
    a_b(t_E)\to a_0 \quad (t_E\to\pm\infty).
\end{align}
We also denote the time at which $a_b(t_E)=a_1$ as $t_E^0$.
Of course, the arbitrariness of $t_E^0$ is related to the zero mode of the bounce solution.

We can now compute the action of the bounce solution, $S_b$.
We first deal with half of the bounce, until $t_E=t_E^0$, and then double it to get the final result, which becomes (by using \eqref{eq:energyconservation})
\begin{align}
    {S_b}= 2\alpha(D)\int_{-\infty}^{t_E^0} dt_E\, (S_E[a_b]-S_E[a_0])
    =2\alpha(D)\int_{a_0}^{a_1} da\, \sqrt{2(V(a)-V(a_0))},
\end{align}
where $\alpha(D)$ comes from integrating over the spatial coordinate.
For later convenience, we redefine $a\equiv b/\sqrt{\abs{g}}$ (Remember that $g<0$.), so that
\begin{align}
    V(b)= \frac{1}{\abs{g}}\left[\frac{\gamma^2}{2b^2}+\frac{m^2 b^2}{2}-\frac{b^4}{4}\right]\equiv \frac{1}{{\abs{g}}}W(b),
    \label{eq:rescaledpotential}
\end{align}
where, as already defined, $\gamma\equiv -g\rho$.
By using these, we have
\begin{align}
    {S_b}=\frac{2\sqrt{2}\alpha(D)}{\abs{g}}\int_{b_0}^{b_1} db\, \sqrt{W(b)-W(b_0)}\equiv \frac{2\sqrt{2}\alpha(D)}{\abs{g}}s_b(\gamma),
    \label{eq:actionofthebounce}
\end{align}
where $a_{0,1}\equiv b_{0,1}/\sqrt{\abs{g}}$.
We conduct the actual computation of this integral in the next subsection.

\subsection{Imaginary part of the operator dimension}

\begin{figure}[tbp]
    \begin{center}
        \begin{overpic}[width=0.55\columnwidth]{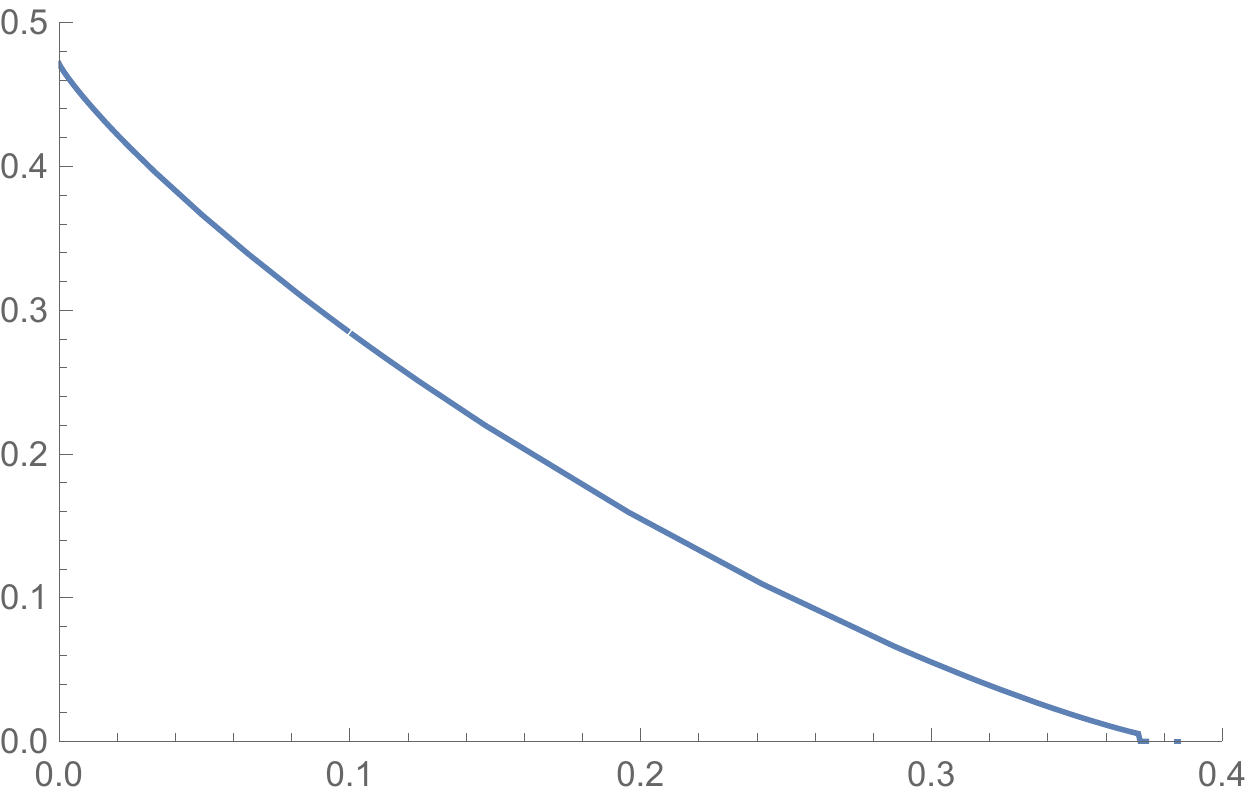}
        \put(0,65){\footnotesize $s_b(\gamma)$}
        \put(100,5){\footnotesize $\gamma$}
        \put(90,2){\footnotesize $\gamma_0$}
        \put(10,70){\footnotesize The action of the bounce as a function of $\gamma$}
        \end{overpic}
    \end{center}
    \caption{The bounce action $s_b$ as a function of $\gamma=g\rho$. It is a decreasing function and it hits zero when $\gamma=\gamma_0$.}
    \label{fig:bounceaction}
\end{figure}
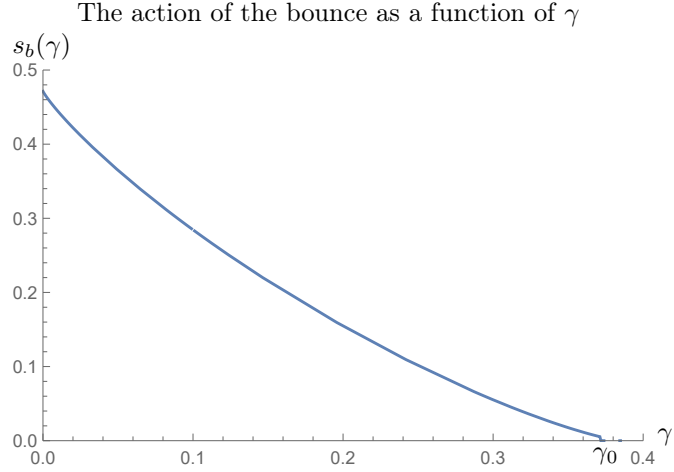

Let us study the bounce action $s_b(\gamma)$ in more detail.
First of all, although there is an analytic expression for $s_b(\gamma)$, the expression is very complicated and we will not show it here.
We instead computed it numerically and showed it in Fig. \ref{fig:bounceaction}.

It is also easy to compute its value at some special values of $\gamma$.
At $\gamma=0$, the value of the action is shown to be
\begin{align}
    s_b(0)=\int_{0}^{\sqrt{2}m} db\, \sqrt{\frac{m^2 b^2}{2}-\frac{b^4}{4}}=\frac{\sqrt{2}}{3}m^3
    \label{eq:epsilonq0}
\end{align}
On the other hand, since $b_0=b_1$ at $\gamma=\gamma_0$, we also immediately see that $s_b(\gamma_0)=0$.

For later convenience, we will define a new object as
\begin{align}
    s_b(\gamma)\equiv \frac{\sqrt{2}}{3}F(\gamma)
\end{align}
so that $F(0)=1$.
In this notation, the action of the bounce becomes
\begin{align}
    S_b=\frac{4\alpha(D)}{3\abs{g}}F(\gamma)
\end{align}

Now that we have computed the action of the bounce, by using the dilute gas approximation it is immediate to see that it contributes to the operator dimension at charge $Q$, as
\begin{align}
    \Delta(Q)&=\Delta_{\rm pert}(Q)+iJ_{b} \exp \left({-S_b}\right)+\cdots\\
    &=\Delta_{\rm pert}(Q)+iJ_{b} \exp \left(-\frac{4\alpha(D)}{3\abs{g}}F(\gamma)\right)+\cdots.
\end{align}
Here, $J_{b}$ is a coefficient which comes from the one-loop determinant around the bounce, and the imaginary number in front comes from a negative eigenvalue of the fluctuation mode.

Borrowing the preceding discussion given in Sec. \ref{sec:epsilonQsmall} and in Sec. \ref{sec:epsilonQtransition} to estimate $J_{b}$, the final result for the imaginary part of the lowest operator dimension in the rank-$Q$ traceless symmetric representation becomes
\begin{align}
	\im \Delta(Q)=\pm f_b(\epsilon Q)\sqrt{\frac{2(N+8)}{\pi\epsilon}}\exp \left(-\frac{N+8}{3\epsilon}F(\epsilon Q)\right)+\cdots,
	\label{eq:final}
\end{align}
where $f_b(\epsilon Q)$ is normalised so that $f_b(0)=1$.
Note that we have swapped the variable of the function $F$ from $\gamma$ to $\lambda\equiv \epsilon Q$, so that now we have $F(\lambda=\lambda_0)=0$, where $\lambda_0\equiv \frac{N+8}{6\sqrt{3}}$.
The result is indeed non-perturbative in $\epsilon$ as promised.
Furthermore, if we think about taking the large-$N$ limit, it is non-perturbative in the $1/N$-expansion as well.

\subsubsection{Analysis near $\epsilon Q=0$}
\label{sec:epsilonQsmall}

Let us compute $\im \Delta(Q)$ around $\epsilon Q=0$.
At $\gamma=0$, the problem is nothing but the computation of the decay rate of the quantum mechanics with an inverted quartic double-well potential with action
\begin{align}
    S[a]=\alpha(D)\left[\frac{1}{2}\left(\partial a\right)^2+V(a;\gamma=0)\right].
    \label{eq:action0}
\end{align}
One can redefine $a\equiv \frac{x}{\sqrt{\alpha(D)}}$ to normalise the kinetic term, in which case we rewrite the potential as
\begin{align}
    V(x;\gamma=0)\equiv \frac{m^2}{2}x^2+\frac{g}{4\alpha(D)}x^4.
\end{align}

The decay rate of the vacuum, which we denote $\im E_{\rm gs}$, is already computed in the literature \cite{doi:10.1142/6050} and is known to be
\begin{align}
    i\im E_{\rm gs}=\pm im\sqrt{\frac{8\alpha(D)}{\pi \abs{g}}}\exp\left[-\frac{4\alpha(D)}{3\abs{g}}\right]\times \left[1+O(g)\right].
    \label{eq:unstablegs}
\end{align}
Hereafter we will not write the $O(g)$ corrections explicitly anymore. 
Incidentally, the exponent $\frac{4\alpha(D)}{3g}$ indeed matches $S_b$ at $\gamma=0$ as it should be.
From this, we get
\begin{align}
    J_b(\gamma=0)=\pm m\sqrt{\frac{8\alpha(D)}{\pi \abs{g}}}.
\end{align}

Incidentally, $J_b$ is in general a function of $g$ and $\gamma$.
Since the potential \eqref{eq:rescaledpotential} depends on $g$ as an overall constant, we can see that 
\begin{align}
    J_b(\gamma)\equiv J_b(0)f_b(\gamma)=\pm m\sqrt{\frac{8\alpha(D)}{\pi \abs{g}}}\times f_b(\gamma)=O\left(\abs{g}^{-1/2}\right)
\end{align}
where $f_b(\gamma)$ is some unknown function of $\gamma$ of order unity, with $f_b(0)=1$.
This gives the final result that we presented in \eqref{eq:final}



\subsubsection{Analysis near $\epsilon Q=\lambda_0$}
\label{sec:epsilonQtransition}

We can also compute $\im \Delta(Q)$ around $\epsilon Q=\lambda_0$.
As in \eqref{eq:action0}, our truncated action is
\begin{align}
    S[a]=\alpha(D)\left[\frac{1}{2}\left(\partial a\right)^2+V(a;\gamma)\right].
    \label{eq:action1}
\end{align}
Let us now set $\gamma\equiv \gamma_0-\delta\gamma$.
If we define a new coordinate $y$ by using $b\equiv \sqrt{\frac{2}{3}}+y (\delta\gamma)^{1/2}$, and furthermore set
$z\equiv y+{2^{-1/2}\cdot 3^{-1/4}}$, we have
\begin{align}
    W(z)&={\rm (const.)}
    -2^{3/2}\cdot 3^{-1/2}z^2\left(z-3^{3/4}\cdot 2^{-1/2}\right)\times (\delta\gamma)^{3/2}+O((\delta\gamma)^{2})\\
    W_0(z)&\equiv -2^{3/2}\cdot 3^{-1/2}z^2\left(z-3^{3/4}\cdot 2^{-1/2}\right)
    \equiv -\frac{1}{2}p^2z^2(z-q^2)
\end{align}
In this scaling limit, the truncated action becomes (modulo an addition of an irrelevant constant)
\begin{align}
    S[z]=\alpha(D)\left[\frac{\delta\gamma}{2\abs{g}}\left(\partial z\right)^2+\frac{(\delta\gamma)^{3/2}}{\abs{g}}W_0(z)\right].
\end{align}
Finally normalising the kinetic term, we get
\begin{align}
    S[x]=\frac{1}{2}\left(\partial x\right)^2-\frac{p^2\abs{g}^{1/2}}{2\alpha(D)^{3/2}}x^2\left(x-q^2\sqrt{\frac{\alpha(D)\delta\gamma}{\abs{g}}}\right)
\end{align}
where $x^2\equiv \frac{\alpha(D)\delta\gamma}{\abs{g}}z^2$.

The imaginary part of the ground state energy of such a system is also already computed in the literature \cite{doi:10.1142/6050}.
As the result is nothing but the imaginary part of the operator dimension that we want, we get
\begin{align}
    i \im \Delta(Q;\gamma_0)=\pm i \left(\delta\gamma\right)^{\frac{7}{8}}\sqrt{\frac{C_0}{\abs{g}}}\exp\left[-\frac{C_1\left(\delta\gamma\right)^{\frac{5}{4}}}{\abs{g}}\right],
    \label{eq:singular}
\end{align}
where $C_{0,1}$ are some cumbersome constants irrelevant to the discussion.
This result tells us the behaviour of the function $F(\gamma)$ and $f_b(\gamma)$ near the transition point, $\gamma=\gamma_0$.
The fact that the exponent is a fractional power of $\delta\gamma$ will be important in the next section.

\section{Discussion}
\label{sec:discussion}

There are two special points of interest in the final result given in \eqref{eq:final} -- at $\epsilon Q=0$ and at $\epsilon Q=\lambda_0$.
The former point is connected to the region where $Q$ is fixed, while the latter is connected to the region $\epsilon Q>\lambda_0$, in which the complex saddle-point dominates.
In the middle, the exponents of the non-perturbative correction gets smaller and smaller, and hence the imaginary part of the operator dimension grows as we increase $\epsilon Q$.

\subsection{$\epsilon Q=0$ -- Reproducing the finite-$Q$ result}

Even though our result was derived in the double-scaling limit in which $Q$ is taken to infinity, if we plug in finite $Q$ into the final expression and sum all the relevant contributions, the result should reproduce the finite-$Q$ result.
This means that $\epsilon Q=0$ should be smoothly connected to the regime of finite-$Q$.
For example, doing so will reproduce the real part of the operator dimension of $\phi^a$ at finite-$Q$, computed using Feynman diagrams.


For the non-perturbative contributions, taking $Q$ to be finite means taking $\epsilon Q$ to be of order $O(\epsilon)$.
Now, \eqref{eq:epsilonq0} and Fig. \ref{fig:bounceaction} means that $F(\epsilon Q)=1-\kappa \epsilon Q +O((\epsilon Q)^2)$ for some positive value $\kappa$.
Therefore, we get
\begin{align}
	\im \Delta(Q)=\pm \sqrt{\frac{2(N+8)}{\pi\epsilon}}\exp\left(\frac{\kappa Q(N+8)}{3}\right)\times \exp \left(-\frac{N+8}{3\epsilon}\right)+\cdots
	\label{eq:qfinite}
\end{align}
where we have set $f_b(\epsilon Q)\approx 1$ as the correction to this is not important.

Comparing this to \cite{McKane:1984eqmod}, we see that the result correctly reproduces the exponent of the non-perturbative correction at $Q=1$.
The coefficient in front is, however, subject to other corrections.
For example, there could be other class of corrections whose exponent is the same as above at $\epsilon Q=0$, but net larger for $\epsilon Q>0$.
All in all, we believe that only the exponent should be correctly reproduces at finite $Q$ from the double-scaling limit, which we succeeded in.
However, as a general trend, we can imagine that the coefficient gets bigger as we increase $Q$.
It would be interesting to check this using conventional analyses.

There is one additional comment regarding this.
In \cite{Giombi:2020enj}, it was noted that the behaviour of the exponent of the non-perturbative correction at $Q=1$ and at large-$N$ is similar to that of $Q/N$ at which the real saddle-point ceases to exist, as functions of $D$.
Incidentally, the latter is an analogue of $\lambda_0$ in the $D=4+\epsilon$ case.
Since the former is the value of the bounce action at $Q/N=0$ and the latter is the value of $Q/N$ at which the bounce action vanishes, there is no wonder that these two are qualitatively similar.
It would be also interesting to conduct the similar computation as ours at large-$N$ to make this connection more concrete.

\subsection{$\epsilon Q=\lambda_0$ -- Condensation of bounces}
\label{sec:instantoninduced}

Even more surprising is the point at which $\epsilon Q=\lambda_0$.
Here, the bounce action vanishes and the leading non-perturbative correction becomes of order one.
It means that the dilute-gas approximation we have used is no longer valid at $\epsilon Q=\lambda_0$, and multi-bounce corrections become equally important.
This phenomenon is called the condensation of bounces, and an analogous phenomenon is observed at the Gross-Witten-Wadia transition point \cite{Wadia:2012fr,Wadia:1980cp,Gross:1980he,Gross:1994mr,Gross:1994ub,Witten:1991we}, called the condensation of instantons (where the instanton action becomes zero at the transition point).

As discussed in \cite{Marino:2008ya,Ahmed:2017lhl,Ahmed:2018gbt}, the condensation of bounces is closely related to the fact that $\Delta(Q;\lambda)$ experiences a phase transition at $\lambda=\lambda_0$ in the strict $Q\to\infty$ limit (See Sec. \ref{sec:phasetransition}).
The argument is that such a condensation prohibits $\Delta(Q)$ from having a phase transition at finite but large $Q$.\footnote{Since $\Delta(Q)$ is nothing but the ground state energy on a finite volume space, the general principle prohibits it from having a phase transition at large but finite $Q$.}
As the non-perturbative correction can be as big as the perturbative one when the condensation happens, they can conspire to smooth out the phase transition to a crossover at finite but large $Q$.

This mechanism can be seen more concretely by using a new double-scaling limit.
This will mimic the similar double-scaling limit used in the context of matrix models or topological strings \cite{Marino:2008ya,Marino:2008vx,Marino:2009dp,Ahmed:2017lhl,Ahmed:2018gbt}.
Let us first recall the form of $\Delta(Q,\gamma)$ near $\gamma_0$, with $\gamma=\gamma_0-\delta\gamma$,
\begin{align}
    \Delta(Q,\gamma)&=\Delta_{\rm pert}(Q,\gamma)+\Delta_{\rm np}(Q,\gamma)\\
    \Delta_{\rm pert}(Q,\gamma)&=\frac{1}{\epsilon} F_0(\gamma)+F_1(\gamma)+\epsilon F_2(\gamma)\cdots\\
    \Delta_{\rm np}(Q,\gamma)&=i\left(\delta\gamma\right)^{\frac{7}{8}}\sqrt{\frac{C_0}{\abs{g}}}\exp\left[-\frac{C_1\left(\delta\gamma\right)^{\frac{5}{4}}}{\abs{g}}\right]+\text{(multi-bounce)}
\end{align}
Since we would like to fix the exponent of the non-perturbative correction, the new double-scaling limit becomes which is to take $\gamma\to\gamma_0$ and $Q\to\infty$, while fixing
\begin{align}
    \gamma\to\gamma_0,\quad Q\to\infty\quad  \epsilon (\delta\gamma)^{5/4}\equiv \kappa^{5/4}=\text{(fixed)}.
\end{align}
Incidentally, this double-scaling limit, with $5/4$ being precisely the exponent, is also seen in the double-scaling limit of the cubic matrix model.
This is no surprise as the form of the potential $V(a)$ can be approximated by a cubic function when $\gamma\approx\gamma_0$, as discussed in \ref{sec:epsilonQtransition}.

In this double-scaling limit, the reorganisation of expansions in terms of $\epsilon$ happens.
The new expansion parameter becomes $\epsilon^{4/5}$ as we have
\begin{align}
    \frac{1}{\epsilon}F_0(\epsilon,\kappa)=\frac{1}{\epsilon}\left(\frac{1}{3}-\frac{\kappa}{\sqrt{3}}\epsilon^{4/5}-\frac{2\kappa^{3/2}}{3^{5/4}}\epsilon^{6/5}+\frac{\kappa^2}{12}\epsilon^{8/5}+\cdots\right).
    \label{eq:aaa}
\end{align}
We also have
\begin{align}
    \Delta_{\rm np}(\epsilon,\kappa)\propto i\epsilon^{1/5}\kappa^{7/8}\exp\left({-\kappa^{5/4}}\right)+\cdots,
\end{align}
and this means that possible singularities physical quantities at $\kappa=0$ due to $\kappa^{3/2}$ in \eqref{eq:aaa} can get cancelled by the non-perturbative corrections, as they are of the same order, $O(\epsilon^{1/5})$.
We leave more detailed discussions of this new double-scaling limit, detailed mechanism in which the phase transition becomes a crossover, or its possible relation to matrix models to future work.

Before concluding the section, we point out the difference of the current situation with that of the Wilson-Fisher fixed point in $D=4-\epsilon$.
In $D=4-\epsilon$, the double expansion is believed not to have non-perturbative corrections, so that there exists no phase transitions as we move from small $\epsilon Q$ to large $\epsilon Q$.
This is in spite of the fact that the convenient description of the theory changes along the way also in $D=4-\epsilon$.
However, it would be interesting to check if there are really no analogous phenomena in $D=4-\epsilon$.

\section{Conclusions and Outlook}
\label{sec:conclusion}

In this paper, we computed the lowest operator dimension in rank-$Q$ traceless symmetric representation of $O(N)$ Wilson-Fisher fixed point in $D=4+\epsilon$.
In particular, we computed, for the first time, the non-perturbative imaginary correction to the operator dimension in the double-scaling limit where $\epsilon Q$ is fixed.
The correction was found by noticing that there is a time-dependent saddle-point, called the bounce, in the Euclidean path integral representation of the projected partition function.
The result of the leading non-perturbative correction turned out to be
\begin{align}
	i\Delta_{\rm np}^{\rm leading}(Q)=\pm if_b(\epsilon Q)\sqrt{\frac{2(N+8)}{\pi\epsilon}}\exp \left(-\frac{N+8}{3\epsilon}F(\epsilon Q)\right).
\end{align}

By using this result, we looked closely at two points of interest, $\epsilon Q=0$ and $\epsilon Q=\lambda_0$.
First, at $\epsilon Q=0$, we saw that the exponent of the non-perturbative correction reproduces the computation of \cite{McKane:1984eqmod} at $Q=1$.
This is interesting since it is the first time where one can demonstrate that the small $\epsilon Q$ expansion is connected to the finite $Q$ region, even in the presence of non-perturbative corrections.
Note that in \cite{Sharon:2020mjs}, the computation in the double-scaling limit did not reproduce the result at finite $Q$, possibly due to the instanton-like corrections.
It would be also interesting to demonstrate the connection between the small $\epsilon Q$ limit and the finite $Q$ region by finding the non-perturbative corrections in their model.

Second, we found a second order phase transition at $\epsilon Q=\lambda_0$ for $\Delta_{\rm pert}(Q)$ in the strict $Q\to\infty$ limit.
This is the point where the real solution to the saddle-point equation ceases to exist.
Even though the model itself is very different, this was analogous to Gross-Witten-Wadia phase transition in that the condensation of bounces occurs at the transition point.
We also computed how fast the bounce action vanishes near the transition point, and argued that they are responsible for cancelling the jumps in the $n$-th derivatives of $\Delta_{\rm pert}(Q)$ in the strict $Q\to\infty$ limit.
We also found a new double-scaling limit which is suited for analysing near the transition, and pointed out its similarity to matrix models.

There are number of interesting future directions which have not been mentioned  in this paper yet.
First, one should try doing the same analysis in $D=6-\epsilon$ or at large-$N$ in general $4<D<6$.
In particular, by doing the large-$N$ expansion, one should be able to compute the value of $Q/N$ at the similar transition point as a function of $D$.
This will be a powerful check of our current result.
It would be also interesting to extend this to other models with a possible double-scaling structure, such as various SUSY theories \cite{Bourget:2018obm,Hellerman:2018xpi,Grassi:2019txd,Sharon:2020mjs,Hellerman:2021yqz,Hellerman:2021duh} or Chern-Simons-matter theories \cite{Watanabe:2019adh,Cuomo:2021qws}. 
One could also look for other double-scaling limit in systems with boundaries or defects \cite{Hoyos:2018jky,Beccaria:2018owt,Billo:2019fbi,Bianchi:2019dlw,Galvagno:2020imh,Giombi:2020amn,Cuomo:2021cnb,Cuomo:2022xgw,Rodriguez-Gomez:2022gbz,Giombi:2022anm}, or with non-relativistic conformal symmetry \cite{Kravec:2018qnu,Favrod:2018xov,Orlando:2020idm,Hellerman:2020eff,Pellizzani:2021hzx,Hellerman:2021qzz}

Second, it is important to do the similar analysis at large $\epsilon Q$.
Even though we saw the divergence in the second derivative of $\Delta(Q)$ in terms of $\lambda$, the analysis is in fact not complete unless we do the same analysis from large $\epsilon Q$.
This region is more interesting but complicated, firstly because the bounce solution will turn into an instanton solution but in the complex plane, connecting two vacua which are complex conjugate to each other.
Secondly, there could be much bigger non-perturbative corrections of order $O(\exp(-\sqrt{\epsilon Q}))$ to the one-loop correction to the static saddle-point already.\footnote{The square root in the exponent of the non-perturbative correction is something which is typical in string theory or in matrix models, as opposed to ordinary QFTs. This might be hinting the connection of the large charge expansion to matrix models or to string theory. This might have something to do with the effective theory of long strings, as in \cite{Hellerman:2013kba,Hellerman:2014cba,Aharony:2013ipa}. We thank Domenico Orlando for discussions.}
As discussed in \cite{Dondi:2021buw,Antipin:2022dsm}, this comes from the worldline instanton of the massive particle on top of the saddle-point.\footnote{In \cite{Antipin:2022dsm} it is argued that the one-loop determinant is convergent around $D=4$. This is indeed correct, but there still exist non-perturbative corrections from the worldline instanton.}

Additionally, even though we found a phase transition, it does not tell us how the degrees of freedom reorganises itself to give a large imaginary part of $\Delta(Q)$ there.
It is indeed an intriguing question where such a large imaginary part comes from in the language of the ordinary perturbation theory.\footnote{It could be that the non-perturbative imaginary part we computed in the present paper somehow reorganises itself into the large imaginary part at large $\epsilon Q$. However, since the imaginary part at large $\epsilon Q$ comes from what was $\Delta_{\rm pert}$ at small $\epsilon Q$, it would be very mysterious and begs for more explanation if this happens.}
Answering this question might be one step towards understanding the similar question in $D=4-\epsilon$, or more in general, the exponentiation of multiparticle amplitudes \cite{Libanov:1994ug,Libanov:1995gh,Son:1995wz,Bezrukov:1995qh}.

Furthermore, the step next would be to study a different double-scaling limit which zooms into the transition point.
This analysis has been conducted already for the Gross-Witten-Wadia model, where one sees that the singularity gets resolved, reflecting the physics of large but finite $N$ \cite{Alvarez-Gaume:2006fwd}.
Through this analysis, we would be able to understand more the relation between the double-scaling limit and the finite $Q$ or finite $\epsilon$ physics.

It can also be an interesting challenge to look for or to disprove non-perturbative corrections in $D=4-\epsilon$.
Usually, it is believed that there are no such corrections in $D=4-\epsilon$, since the leading singularity in the Borel plane in the $\epsilon$-expansion lies in the negative real axis \cite{Dunne:2021lie}.
However, in the double-scaling limit, we can think about a bounce solution, which, this time, goes off to the complex plane and then comes back to the real axis.
These types of complex non-perturbative objects may or may not contribute to the physical quantity as discussed in \cite{Behtash:2015zha,Behtash:2015loa,Buividovich:2015oju,Dunne:2016jsr}, and it would be interesting to check either way.

Thinking about the resurgence structure of the current double expansion is also interesting.
In some cases, non-perturbative corrections can be reproduced from the perturbative expansion \cite{Dunne:2013ada,Dunne:2014bca,Misumi:2015dua,Dunne:2016qix,Gahramanov:2016xjj}.
This might mean that by studying $\Delta_{\rm pert}$ carefully, we might be able to extract information about the non-perturbative correction that we computed.
What might be related is the concavity of the operator dimension as a function of $Q$ at small $\epsilon Q$.
Such a concavity is conjectured to indicate some sickness of the theory (\it e.g., \rm non-unitarity) in relation to the weak gravity conjecture \cite{Aharony:2021mpc,Antipin:2021rsh} (For a review of weak gravity conjecture in general see \cite{Harlow:2022gzl}.)
As discussed in \cite{Moser:2021bes,Orlando:2021usz}, this inevitably leads to complex operator dimensions at large $\epsilon Q$.
However, the theory of resurgence might tell us how the sickness (in this the imaginary part in the operator dimension) comes about already at small $\epsilon Q$.


More generally, this double-scaling limit can be thought of as an example where a large particle number changes the qualitative behaviour of the system, even if it is weakly-coupled.
This phenomenon is in fact ubiquitous in physics.
For example, Gross-Witten-Wadia transition, which we already saw is similar to our transition, is holographically dual to the string-Black Hole transition \cite{Susskind:1993ws,Horowitz:1996nw,Sen:1995in,Alvarez-Gaume:2005dvb,Alvarez-Gaume:2005dvb}.
Now, this transition can happen entirely at weak-coupling, but as we increase the temperature the excitation level of a string increases, and after a certain temperature the state looks like a small Black Hole.
It would be therefore interesting to think about the connection between the double-scaling limit of the large charge expansion, and the string-Black Hole transition.
More concretely, it would be nice to come up with a finite-temperature version of our story, and to make parallel comparison of the semi-classical saddle with the semi-classical gravity configuration governing the Horowitz-Polchinski solution.

It would also be interesting to think about the holographic dual of our transition in a direct way.
Our CFT picture is that there are $Q$ weakly interacting particles, and as we increase the particle number, at some threshold particle density the transition happens and the state starts looking like a superfluid, homogeneously distributed on the spatial slice. 
Therefore, a very qualitative holographic picture of the transition should be that these $Q$ particles near the boundary of the $AdS$ space gets attracted more to the centre of it, finally reaching some semi-classical object at the centre of $AdS$.\footnote{Unfortunately, there is still no consensus as to what such a semi-classical object in gravity should look like, although there are several proposals so far \cite{Loukas:2018zjh,delaFuente:2020yua,Liu:2020uaz,Guo:2020bqz}.}


Finally, as a distant goal, it might be worthwhile to understand the relation to matrix models.
First of all, at least superficially by comparing the series expansion, we can see that the 't Hooft expansion and our double-scaling expansion is parallel.
For example, in the case of ABJM theory \cite{Aharony:2008ug}, the rank of the gauge group $N$ can be identified with $Q$, whereas the level $k$ can be with $1/\epsilon$.
In this analogy, the large charge regime $Q\gg 1$ and fixed $\epsilon$ corresponds to the M-theory regime.

In fact, the relation might be a little more than just being superficial.
In the $M$-theory regime, it is known that there is a description in terms of Fermi gases \cite{Marino:2011eh}.
Here, $k$ is proportional to the Planck constant of the Fermi gas system, whereas $N$ corresponds to the number of particles, which is somehow close enough to the charge.
Furthermore, the relation between the matrix model (or its Fermi gas picture) and the large charge expansion is already pointed out in the context of $\mathcal{N}=2$ SQCD in $D=4$ \cite{Grassi:2019txd,Bissi:2021rei}.
A similar phenomenon is also observed in the context of Wilson loops in $\mathcal{N}=4$ SYM \cite{Giombi:2021zfb,Giombi:2022anm}.
It would be of great use if we can come up with the large charge effective action corresponding to the Fermi gas system, and even better, to the $M$-theory regime of ABJM theory.

As another supporting evidence, we have also found in this paper that the new double-scaling limit near the phase transition point is parallel between the large charge expansion and the matrix model.
The analogy will be more convincing if we study the limit $\kappa\to\infty$, which we leave for future work.
This limit should be connected to the regime of large $Q$ and fixed $\epsilon Q\ll 1$, and it would be interesting to first check this.
Additionally, it would be interesting to check what the expansion parameters are for both perturbative and non-perturbative parts.
It was argued in the context of the matrix model that the expansion parameter of the non-perturbative part is square root of that of the perturbative part \cite{Aniceto:2011nu}. 
This is because the former comes from the open string coupling while the latter, the closed string coupling.
It would reinforce the analogy and possibly help find the matrix model dual of our theory if we find the same phenomena in our case as well.




%
%
%

\section*{Acknowledgement}
The author thanks Ohad Mamroud and Adar Sharon for collaborations at early stages of the work, as well as stimulating discussions and useful advice throughout.
We are also grateful to Ofer Aharony, Simeon Hellerman, Domenico Orlando and Susanne Reffert for valuable discussions and the thorough reading of the manuscript.
This work is supported by the Foreign Postdoctoral Fellowship Program of the Israel Academy of Sciences and Humanities, by Israel Science Foundation center for excellence grant (grant number 2289/18), and by the German Research Foundation through a German-Israeli Project Cooperation (DIP) grant ``Holography and the Swampland''.

\bibliographystyle{JHEP}
\bibliography{condmat,main}

\end{document}